\documentclass[prd,aps,showpacs,tightenlines,nofootinbib,preprint,preprintnumbers]{revtex4}
\usepackage{graphicx}
\usepackage{amssymb}

\newcommand{\bear}{\begin{array}}  \newcommand{\eear}{\end{array}}
\newcommand{\bea}{\begin{eqnarray}}  \newcommand{\eea}{\end{eqnarray}}
\newcommand{\beq}{\begin{equation}}  \newcommand{\eeq}{\end{equation}}
\newcommand{\bef}{\begin{figure}}  \newcommand{\eef}{\end{figure}}
\newcommand{\bec}{\begin{center}}  \newcommand{\eec}{\end{center}}
\newcommand{\non}{\nonumber}

\newcommand{\del}{\partial}

\newcommand{\la}{\left\langle} \newcommand{\ra}{\right\rangle}

\begin{document}

\title{$D$-term Instability in Flat Directions 
and its Cosmological Implication}
\author{Masahiro Kawasaki}
\author{Fuminobu Takahashi}
\affiliation{Research Center for the Early Universe, University 
  of Tokyo, Bunkyo-ku, Tokyo 113-0033, Japan}
 \affiliation{
  Institute for Cosmic Ray Research, University of Tokyo,
  Kashiwa, Chiba 277-8582, Japan}
\date{\today}

\begin{abstract}
We study dynamics of flat directions in the minimal
supersymmetric standard model taking account of its constituent 
fields. 
It is found that there exist new instabilities due to the 
D-term potential and the nature of these instabilities 
depends on the eccentricity of the orbit. For a roughly
circular orbit, it is similar to  narrow-band resonance
described by the Mathieu equation. For an  elliptic
orbit,  the instabilities exhibit tachyonic nature. 
In the narrow-band regime, we show that the existence of the 
instabilities is crucial to the formation of $Q$-balls. 
The $Q$-ball formation proceeds through  quasi-stable state called  ``$I$-ball''.
The transition from $I$-balls to $Q$-balls is made efficient
by the $D$-term instability and during this process some fraction
of the charge of the $Q$-ball is emitted. 
This discovery may revive the scenario which relates the baryon 
number and dark matter density of the universe. 
Furthermore, the tachyonic $D$-term instability causes the drastic decay of the
flat direction moving in an orbit with relatively large eccentricity.
Thus the evolution of flat directions is completely
altered by the appearance of this instability. 
 
\end{abstract}
\pacs{98.80.Cq \hspace{8.0cm} }
\maketitle

\section{introduction} 
\label{sec:intro}
The scalar potential in the minimal supersymmetric standard 
model (MSSM) has flat directions along which there is no
classical potential in supersymmetric (SUSY) limit. 
In the real world, the potential along the flat directions
is not completely flat because SUSY breaking induces soft
masses of the order $\sim O(100)$~GeV.  
Such flat directions  play important roles in cosmology.   
In the inflationary universe some fields corresponding to
flat directions have large field values during inflation, and 
start oscillation after their mass becomes comparable to the
Hubble parameter. 
The flat directions generally consist of squarks, sleptons 
and Higgs, and hence have baryon and/or lepton numbers. 
Thus, the oscillation of the flat directions and subsequent decay
can produce large baryon and/or lepton numbers in the universe
if some baryon(lepton) number violating terms exist in the potential.
This mechanism was discovered by 
Affleck and Dine~\cite{Affleck:1984fy} and it has been
one of promising scenarios for explaining the baryon asymmetry 
of the present universe. Hereafter we call scalar fields 
corresponding to flat directions Affleck-Dine (AD) fields. 

Recently, it was found that  condensates of some AD fields
evolve into non-topological solitons called 
$Q$-balls~\cite{Coleman:1985ki} through strong instabilities 
during oscillation~\cite{Kusenko:1997si,Enqvist:1998en}.
In fact, numerical simulations showed that initial tiny
fluctuations of the AD field rapidly grow and deform 
into large $Q$-balls~\cite{Kasuya:1999wu,Kasuya:2000wx}. 
The existence of $Q$-balls complicates dynamics of the AD field
and significantly changes the scenario of the 
AD baryogenesis~\cite{Kusenko:1997si,Enqvist:1998en,
Kasuya:2000sc,Kasuya:2001hg}. 
$Q$-balls also provide several interesting possibilities. 
For example, $Q$-balls can account for the dark matter and baryon
asymmetry of the universe simultaneously~\cite{Kasuya:2000sc,
Kasuya:2001hg}, or they may produce very large lepton 
asymmetry~\cite{Kawasaki:2002hq,Ichikawa:2004pb}. 

So far, dynamics of  flat directions has been studied
with use of ``single-field parametrization'', i.e. a flat direction
is described by one complex scalar field. 
In fact, however, the flat direction is actually composed 
of several scalar fields. When one only considers the dynamics 
of a homogeneous AD field, the single-field parametrization 
may be good enough.
However, we now know that the fluctuations of the AD field is
crucial to $Q$-ball formation. Since the field fluctuations 
increase the number of degrees of freedom in the dynamical system, 
the single-field parametrization is not always adequate to
describe the evolution of the fluctuations and multi-field 
analysis is necessary. 
In the previous paper~\cite{Kawasaki:2004th}, 
we demonstrated the importance of
the multi-field analysis when one of component fields that form  
$Q$-balls has a large decay rate.

In this paper, we study the dynamics 
of the AD field taking account of its constituent fields.
We find that there exist instabilities due to the $D$-term 
potential and the nature of these instabilities is determined by
the eccentricity of the orbit. For a roughly circular orbit, it is
described by narrow-band resonance. However, as the eccentricity increases
the instability exhibits another character: tachyonic instability.

In the former case, since the instability is the narrow-band resonance,  
its effect on the cosmological evolution of the homogeneous mode of 
the AD field is not so important. 
However, the existence of the instabilities
is crucial to the formation of $Q$-balls. 
In general, in gravity mediated SUSY breaking models,
the $Q$-ball formation is complete through quasi-stable
state  ``$I$-balls''~\cite{Kasuya:2002zs} inside which the 
AD field rotates in an elliptical orbit. 
It is found that the D-term instability quickly makes the orbit of 
the AD field circular and a $Q$-ball configuration is 
reached.  
During this process some fraction of the charge confined 
in the $Q$-ball is emitted, which is different from the previous 
result obtained with use of the single-field parametrization. 
As a result  the baryon density of the universe is accounted for
by both $Q$-ball and residual baryon emitted during the $Q$-ball formation.
On the other hand the dark matter is explained by the lightest
SUSY particle (LSP) produced in the $Q$-ball decay. 
This may revive the scenario which relates the baryon 
number and dark matter density of the universe. 

Meanwhile,  if the instability becomes tachyonic,  its effect on
the cosmological evolution of the AD field is no longer negligible.
Irrespective of whether $Q$-balls are formed or not, the homogenous
motion always suffers from  rapid grow of the instabilities. 
Therefore  the AD field moving in an orbit with large eccentricity 
decays soon after it begins to oscillate.
We expect that the similar instabilities will exist in many other cases.
For instance, the instabilities might change the reheating processes of some
supersymmetric inflation models. 

In the next section, we investigate the D-term instability, and
the effect on $Q$-ball formation is studied in Sec.~\ref{sec:qb_form}.
Sec.~\ref{sec:tachyonic} is devoted to the discussion on the fate of
the flat directions under the effect of 
the tachyonic instability and 
finally we give discussions and conclusions
in Sec.~\ref{sec:discuss}.

\section{$D$-term instability} 
\label{sec:dterm_inst}
In this section, first we explain how flat directions are expressed,
paying particular attention to the relationship between single-field and multi-field
parametrizations. The former is usually adopted in many
literature, since it well describes the collective motion of the constituent
fields. Interestingly however, for instance, if some of the constituent fields 
decay into lighter species, such simple description
fails~\cite{Kawasaki:2004th}. Furthermore, the $D$-term
instability appears only in the multi-field parametrization, which
enables us to explicitly deal with the $D$-term potential.
After reviewing the flat direction, we will investigate the instability in detail.

\subsection{Flat directions}
In the MSSM, there are many flat directions along which both the
$D$-term and $F$-term potentials vanish at the classical level.  The
$D$-flat direction is labeled by a holomorphic gauge-invariant
monomial, $X$. Most of the $D$-flat directions can be also $F$-flat,
simply due to a generation structure of the quark and lepton sector.
Since it is easy to satisfy the $F$-flat condition, let us concentrate
on the $D$-flat condition in the following. The $D$-flat direction,
$X$, can be expressed as
\beq
\label{eq:monomial}
X \equiv \prod_{i}^{N} \Phi_i,
\eeq
where $N$ superfields $\left\{ \Phi_i \right\}$ constitute the flat
direction $X$, and we have suppressed the gauge and family indices
with  understanding that the latin letter $i$ contains all the
information to label those constituents.  When $X$ has a nonzero
expectation value, each constituent field also takes a nonzero
expectation value
\begin{equation}
\label{eq:chivev}
\la \Phi_i \ra = \frac{\phi_i}{\sqrt{2}} e^{i \theta_i}.
\end{equation}
Here each $\phi_i/\sqrt{2}$ is the absolute value of the expectation
value $\la \Phi_i \ra$, and is related to every other due to the
$D$-flat conditions:
\beq
\label{eq:D-flat}
V_D(\left\{\Phi_i\right\}) \equiv
\frac{1}{2} \sum_A \left(
\sum_{ij} \Phi_i^* (T_A)_{ij} \Phi_j\right)^2 = 0,
\eeq
where $T_A$ are hermitian matrices representing the generators of the
gauge algebra, and they are labeled with $A$. This condition can be
usually satisfied if we take all $\phi_i$ equal: $\phi_i \equiv
\phi/\sqrt{N}$. Note that the $D$-flat condition dictates that all the
amplitudes be equal, while the phases, $\theta_i$, remain to be
arbitrary. In fact, it is necessary to know the pattern of baryon and
lepton symmetry breaking, in order to identify the Nambu-Goldstone
(NG) boson relevant for the AD mechanism. The situation becomes simple
if the spontaneously broken symmetry coincides with the explicitly
violated one. This is the case if the A-term is given by some powers
of $X$. Then all $\theta_i$ become equivalent, and the NG boson can be
identified with the average of $\theta_i$~\cite{Takahashi:2003db}:
\beq
\theta \equiv \frac{1}{N} \sum_i \theta_i,
\eeq
where $\theta$ is orthogonal to NG modes corresponding to the
spontaneously broken $U(1)$ gauge symmetries: the 
linear combination of left isospin $T_{L3}$ and weak hypercharge $Y$.
This is because 
$\sum_i  T_{L3, i}=\sum_i  Y_i=0$
by definition of the $D$-flat direction.
Thus, the dynamics of the flat direction, $X$, is described by
one complex scalar field, $\Phi$:
\beq
\Phi \equiv \frac{\phi}{\sqrt{2}} e^{i \theta}.
\eeq
Note that $\Phi$ is canonically normalized. The single-field parametrization is so useful in usual situations that we almost forget that the flat direction is
actually composed of multiple fields. However, one must keep in mind that this parametrization is just an
approximation which becomes exact only in the limit of $D$-flatness, 
and that it describes only the collective motion of the constituent fields.
The smallness of the $D$-term potential does not necessarily 
guarantee that its effect on the dynamics of flat direction as well is negligibly small. This is actually what we are going to show in this paper.

Flat directions are lifted by the SUSY breaking effect and
non-renormalizable terms. 
To be concrete, let us adopt the gravity-mediated SUSY breaking model. The 
flat direction is then lifted as
\bea
\label{eq:potential}
V_s(\left\{\Phi_i\right\}) &=& \sum_i m_{i}^2 |\Phi_i|^2 \left(
1+K_i \log\left(\frac{|\Phi_i|^2}{M_*^2}\right)\right)
\eea
where $m_i$ is a soft mass, $K_i$ a coefficient of the one-loop
correction, $M_*$ the renormalization scale to define the mass.  This
potential reduces to the following in the single-field representation,
\bea
\label{eq:phi_pote}
V_s(\Phi) &=&  m_{\Phi}^2 |\Phi|^2 \left(
1+K \log\left(\frac{|\Phi|^2}{N M_*^2}\right)\right),
\eea
where $m_\Phi$ and $K$ are defined as
\beq
m_\Phi^2 =\frac{1}{N} \sum_i m_i^2,~~~K= \frac{\sum_i m_i^2 K_i}{\sum_i m_i^2}.
\eeq
The absolute value of $K$ is estimated as $|K|=0.01 \sim 0.1$~\cite{Enqvist:1998en}.
Moreover, assuming a nonrenormalizable operator in the
superpotential of the form
\begin{equation}
\label{eq:spnr}
    W_{\rm NR} = \frac{X^k}{k M^{N k-3}},
\end{equation}
the flat direction is further lifted by the potential
\bea
\label{eq:pnr}
 V_{\rm NR}(\left\{\Phi_i\right\})  &=&
    \sum_{i=1}^{N}\frac{ |X|^{2k}}{M^{2 Nk -6}|\Phi_i|^2},
\eea
where
$M$ is a cutoff scale.
In fact, the nonrenormalizable superpotential not only lifts the potential but
also gives the baryon and/or lepton number violating A-term of the
form
\begin{eqnarray}
\label{eq:nrA}
 V_{A}(\left\{\Phi_i\right\}) &=& a_m \frac{m_{3/2}}{k M^{N k -3}} X^k +
    	 {\rm h.c.},
\end{eqnarray}
where $m_{3/2}$ is the gravitino mass, $a_m$ is a complex constant of
order unity, and we assume a vanishing cosmological constant. 

During inflation, the flat direction, $X$, is assumed to take a large
expectation value. After inflation ends, the Hubble
parameter starts to decrease, and becomes comparable to the mass of
$\Phi$ at some point. Then the flat direction starts to oscillate and acquires a finite angular momentum due to the A-term.  Since we are interested in the evolution of the flat direction after it starts to oscillate, we neglect  non-renormalizable terms (including the A-term), and take ellipticity of the trajectory as a free parameter. Instead of eccentricity, however,
we use axial ratio
$\varepsilon$ defined as the ratio of the magnitudes of the major axis and
minor axis of ellipse described by the AD field (see Eq.~(\ref{eq:axial})). For instance,
$\varepsilon=1$ corresponds to a circular orbit, while $\varepsilon=0$ represents a straight-line
motion.
In the following sections we study the evolution of $\{\Phi_i\}$
with the potential $V_D(\{\Phi_i\})+V_s(\{\Phi_i\})$ in the multi-field
parametrization, while the potential is simplified as Eq.~(\ref{eq:phi_pote}) in the
single-field parametrization.

\subsection{Instability associated with $D$-term potential}
Now let us consider dynamics of a flat direction taking account
of the $D$-term potential. As a beginning we would like to clarify a setup
relevant for our discussion.
We take up the simplest possible flat direction composed
of two scalar fields with $U(1)$ gauge symmetry,
$X=\Phi_1(+1)\,\Phi_2(-1)$, but it is trivial to extend our results to
more generic case. The effect of the expanding universe is neglected
for the moment, but we will get back to this point later.

The constituent fields, $\Phi_{1,2} $, obey the equations of
motion
\bea
\label{eq:EOM}
\ddot{\Phi}_i+\frac{\del V(\Phi_1,\Phi_2)}{\del \Phi_i^*} &=&0
,~~(i=1,2)
\eea
with
\bea
\label{eq:potential12}
V(\Phi_1,\Phi_2) &=&  m_{1}^2 |\Phi_1|^2 
+m_{2}^2 |\Phi_2|^2 
+
\frac{g^2}{2} \left(
|\Phi_1|^2- |\Phi_2|^2\right)^2,
\eea
where the overdot denotes the differentiation with respect to time,
$g$ is the $U(1)$ gauge coupling constant,
and we take $K_1=K_2=0$  to concentrate on the instability that
originates from the $D$-term.  Rewriting the equations of motion
in the real and imaginary components:
\beq
\Phi_i \equiv \frac{\phi_{iR}+i\,\phi_{iI}}{\sqrt{2}},
\eeq
we obtain
\bea
\label{eq:real1}
\ddot{\phi} _{1R}+m_1^2 \phi_{1R}
+\frac{g^2}{2}\left(\phi_{1R}^2+\phi_{1I}^2-\phi_{2R}^2-\phi_{2I}^2\right) \phi_{1R}&=&0,\\
\label{eq:real2}
\ddot{\phi}_{2R}+m_2^2 \phi_{2R}
+\frac{g^2}{2}\left(\phi_{2R}^2+\phi_{2I}^2-\phi_{1R}^2-\phi_{1I}^2\right) \phi_{2R}&=&0,
\eea
for the real parts. The equations for the imaginary parts can be obtained by
interchanging the subindex $R$ with $I$.
When the flat direction starts to oscillate, its 
amplitude is usually much larger than $m_\Phi/g$, so that 
the $D$-flat condition: $|\Phi_1|=|\Phi_2|$, is satisfied with good precision. 
With an appropriate phase rotation, the $D$-flat condition can be also expressed as
$\phi_{1R}=\phi_{2R}$ and $\phi_{1I}=\phi_{2I}$.
Thus the homogeneous motion can be described by a single field $\Phi$
rotating in a parabolic potential. 

\begin{figure}
\includegraphics[width=7.5cm]{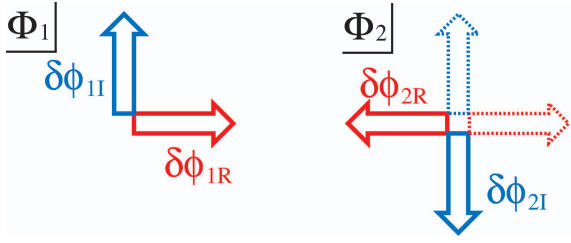}
\caption{
The relative fluctuations between $\Phi_1$ and $\Phi_2$ (solid arrows), which
grows due to the $D$-term instability. The single-field parametrization deals  only with the collective fluctuations shown as dotted arrows. 
}
\label{fig:rel}
\end{figure}

Let us now consider fluctuations of the flat direction moving in an elliptical orbit 
with axial ratio~\footnote{
We do not consider the case that $\varepsilon$ is extremely close to $0$, 
which corresponds to an almost straight-line motion.
This is because other type of instability known as
 ``preheating"~\cite{Kofman:1997yn}
comes into play.
In particular, the instant preheating~\cite{Felder:1998vq,Kasuya:2003iv} could 
proceed if $\varepsilon \lesssim m_{\Phi}/g|\Phi|_{\rm osc}$, where $|\Phi|_{\rm osc}$
represents the amplitude of the flat direction when it begins to oscillate.
Since the flat direction generally acquires a finite angular momentum 
due to the A-term, we can easily avoid such drastic decay. 
}: $0<\varepsilon\leq 1$. First we derive the equations of motion
for linearized fluctuations from 
Eqs.~(\ref{eq:real1}) and (\ref{eq:real2}). Using the $D$-flat condition of the homogeneous
motion, we have
\bea
\label{eq:delR1}
\delta \ddot{\phi}_{1R}+ \left(k^2+m_1^2\right) \delta \phi_{1R}+
g^2 \phi_{1R} \left(\phi_{1R} \delta \phi_{1R}+\phi_{1I} \delta \phi_{1I}
-\phi_{2R} \delta \phi_{2R}-\phi_{2I} \delta \phi_{2I}
\right) &=& 0,\\
\label{eq:delR2}
\delta \ddot{\phi}_{2R}+ \left(k^2+m_2^2\right) \delta \phi_{2R}+
g^2 \phi_{2R} \left(\phi_{2R} \delta \phi_{2R}+\phi_{2I} \delta \phi_{2I}
-\phi_{1R} \delta \phi_{1R}-\phi_{1I} \delta \phi_{1I}
\right) &=& 0,
\eea
and  similar equations for the imaginary parts.  Here and in what follows
we consider fluctuations in momentum space, and $k$ denotes the momentum. 
Further simplification is possible
if $m_1=m_2(=m_\Phi)$, so we assume that this is the case just for simplicity. 
However, note that this assumption is not obligatory, and that it does not 
change generic features of the $D$-term instability. From Eqs.~(\ref{eq:delR1})
and (\ref{eq:delR2}), one can see that $\delta \phi_{1R}+\delta \phi_{2R}$
obeys a simple harmonics equation with a constant frequency. In other words,
$\delta \phi_{1R}+\delta \phi_{2R}$ does not grow at all. Similar arguments
apply to the imaginary parts $\delta \phi_{1I}+\delta \phi_{2I}$ as well. 
Therefore, for a growing mode,
it is a fairly good approximation to take
\bea
\label{eq:rel_app}
\delta \phi_{1R}+ \delta \phi_{2R} = 0,~~~~~\delta \phi_{1I}+ \delta \phi_{2I} = 0.
\eea
Note that this approximation picks out the relative fluctuations between 
the constituent fields $\Phi_1$ and $\Phi_2$, which are disregarded in the single-field parametrization. See Fig.~\ref{fig:rel}.
If all the relative fluctuations
possessed  very heavy masses $\sim g \la \Phi \ra$, the single-field parametrization
might suffice. However, one of the relative fluctuations is actually a massless mode
as shown below, which  experiences the $D$-term instability and  thereby grows rapidly.

\begin{figure}
\includegraphics[width=7.5cm]{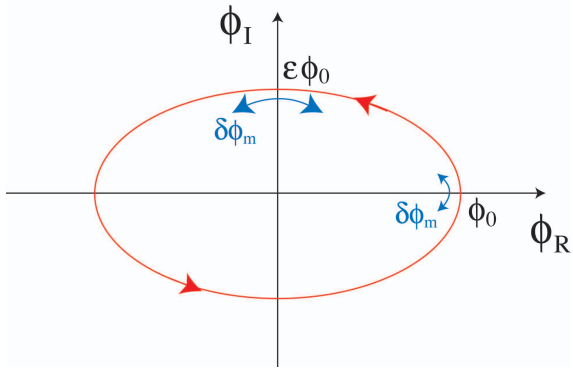}
\caption{
The homogeneous motion and the fluctuation along the elliptical orbit.
}
\label{fig:orbit}
\end{figure}

Hereafter we follow the motion of relative fluctuations $\delta \phi_{R}$ and $\delta \phi_{I}$ defined as
\bea
\delta \phi_{R} &\equiv&\frac{ \delta \phi_{1R}- \delta \phi_{2R}}{2},\non\\
\delta \phi_{I} &\equiv& \frac{\delta \phi_{1I}- \delta \phi_{2I}}{2},
\eea
and drop the subindices $``1"$ and $``2"$ at the homogeneous fields.
The equations of motion for fluctuations $\delta \phi_{R}$ and $\delta \phi_{I}$ are
written in the following form:
\bea
\left(
\bear{c}
\delta \phi_{R}\\
\delta \phi_{I}
\eear
\right)^{\cdot\cdot}
+\omega_k^2 \left(
\bear{c}
\delta \phi_{R}\\
\delta \phi_{I}
\eear
\right) = 0
\eea
with
\bea
\omega_k^2 \equiv \left(
\bear{cc}
k^2+m_\Phi^2+2 g^2 \phi_R^2&2 g^2 \phi_R \phi_I \\
2 g^2 \phi_I \phi_R& k^2+m_\Phi^2+2 g^2 \phi_I^2
\eear
\right).
\eea
In order to diagonalize $\omega_k^2$, we define
$R_k$ and  $\delta \phi_{M,m}$
as
\beq
 \left(
\bear{c}
\delta \phi_{R}\\
\delta \phi_{I}
\eear
\right) = R_k  \left(
\bear{c}
\delta \phi_{M}\\
\delta \phi_{m}
\eear
\right),~~~
R_k\equiv
\frac{1}{\phi_o}\left(
\bear{cc}
\phi_R&-\phi_I\\
\phi_I&\phi_R
\eear
\right),
\eeq
where $\phi_o$ is a constant representing 
the typical amplitude of the flat direction.
The eigenvalues of $\omega_k^2$ corresponding to $\delta \phi_M$ and $\delta \phi_m$ are 
$k^2+m_\Phi^2+2g^2 (\phi_R^2+\phi_I^2)$ and $k^2+m_\Phi^2$, respectively.
Since the frequency of $\delta \phi_M$ is much larger than that of the homogeneous
motion, its evolution is adiabatic, which means that the particle production of 
$\delta \phi_M$ is negligible. The equation of motion for $\delta \phi_m$ is thus
given by
\beq
\label{eq:del_phim}
\delta \ddot{\phi}_{m}-m_\Phi (1-\varepsilon^2)\frac{\sin{2 m_\Phi t}}{\cos^2{m_\Phi t}+\varepsilon^2
\sin^2{m_\Phi t}} \delta \dot{\phi}_m+ k^2 \delta \phi_m = 0,
\eeq
where we have substituted the solution for the homogeneous mode:
\bea
\label{eq:axial}
\phi_R = \phi_o \cos{m_\Phi t},\non\\
\phi_I = \varepsilon \phi_o \sin{m_\Phi t},
\eea
and we have neglected $\delta \dot{\phi}_M$ in Eq.~(\ref{eq:del_phim}) 
since it is rapidly  oscillating. 
$\delta \phi_m$ is 
fluctuation along the elliptical orbit as shown in Fig.~\ref{fig:orbit},
which can be confirmed by noting that Eq.~(\ref{eq:del_phim}) except for the third term
is same as the equation of motion for the angular variable $\theta$ of the homogeneous 
AD field moving in an elliptic orbit with axial ratio $\varepsilon$:
\beq
\ddot{\theta}+2 \frac{\dot{\phi}}{\phi} \dot{\theta} = 0,
\eeq
where $\phi$  and $\theta$ represent the amplitude and angle of $\Phi$ in the
single-field representaion: $\phi^2/2 = \phi_R^2 + \phi_I^2$; $\tan \theta = \phi_I/\phi_R$.
 Using this analogy,
it is easy to see that  Eq.~(\ref{eq:del_phim}) contains instabilities, leading to
the rapid growth of $\delta \phi_m$.
Clearly, the angle $\theta$  increases monotonically as the AD field moves around in its orbit.
In particular, $\theta$ exhibits  step-function-like behavior for $\varepsilon \ll 1$.  
Therefore we can deduce that  $\delta \phi_m$ as well increases monotonically
aside from the case of $\varepsilon = 1$, in which
$\delta \phi_m$ oscillates with the constant angular frequency $k$,
while $\theta$ increases linearly with time. 

To see the band structure of this instability, let us rewrite this equation in terms of
$\tau$ and $\delta \varphi$ 
defined as
\bea
\tau &\equiv& m_\Phi t-\frac{\pi}{2},\\
\delta \varphi &\equiv&  \delta \phi_m \sqrt{1+\varepsilon^2-(1-\varepsilon^2) \cos{2\tau}}.
\eea
Then we obtain the final result:
\beq
\label{eq:fluc_eq}
\delta \varphi^{''}+\left(\frac{k^2}{m_\Phi^2}+ F(\varepsilon, \tau)\right)  \delta \varphi=0
\eeq
with
\beq
F(\varepsilon, \tau)  \equiv \frac{(1-\varepsilon^2)(\sin^4{\tau}-\varepsilon^2 \cos^4{\tau})}{
(\sin^2{\tau}+\varepsilon^2 \cos^2{\tau})^2},
\eeq
where the prime denotes the differentiation with respect to $\tau$.
It is worth noting that the amplitude of the flat direction $\phi_o$ disappears
in the Eqs.~(\ref{eq:del_phim}) and (\ref{eq:fluc_eq}). 
Therefore the instability band and exponential growth rate 
are independent of the amplitude. 

\begin{figure}
\includegraphics[width=7.5cm]{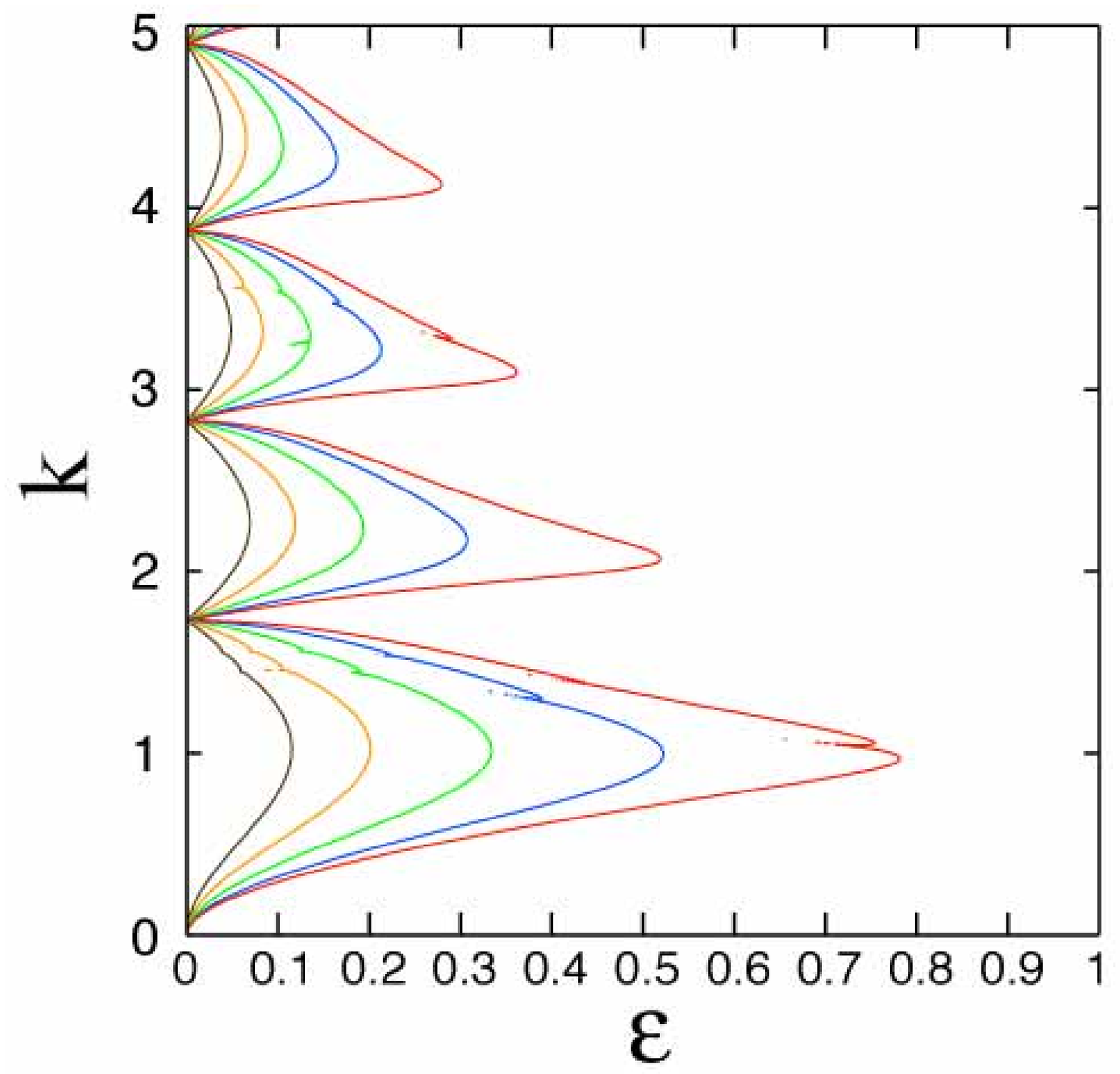}
\includegraphics[width=7.5cm]{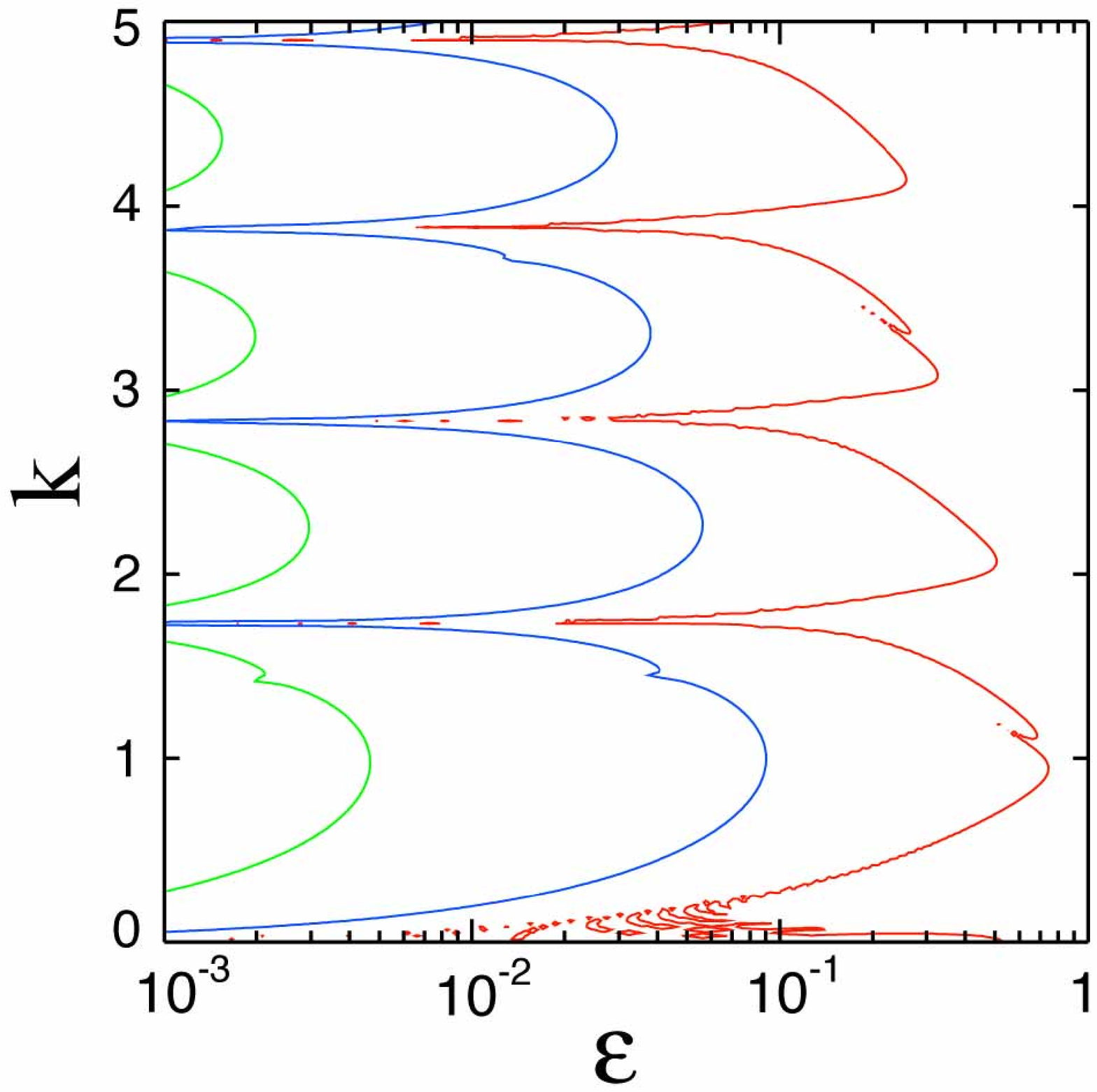}
\caption{
The instability charts of Eq.~(\ref{eq:fluc_eq}). 
The momentum $k$ is normalized by $m_\Phi$.
The left and right panels are
equivalent except for the plotted range of $\varepsilon$.
Each contour line corresponds to $\mu/m_\Phi = 0.1, 0.3, 0.5, 0.7, 0.9$ from
right to left in the left panel, while
$\mu/m_\Phi  =  0.1, 1, 2$ from right to left in the right panel, 
where $\mu$ is an exponential growth rate of fluctuations.
}
\label{fig:ins_chart}
\end{figure}

First let us  take the limit of $\varepsilon \rightarrow 1$ in Eq.~(\ref{eq:fluc_eq}).
It then reduces to the well-known Mathieu equation~\cite{mathieu}:
\beq
\delta \varphi^{''} + (A-2q \cos{2 \tau}) \delta \varphi=0
\eeq
with $A=k^2/m_\Phi^2$ and vanishingly small $q$-parameter $q\simeq1-\varepsilon$.
For $q \ll 1$, the resonance is known to occur only in narrow bands near 
$A = l^2$, $(l=1,2,\dots)$, and the instability vanishes if $q=0$. 
Therefore we expect that the nature of the instabilities  
of Eq.~(\ref{eq:fluc_eq}) is similar to that of the Mathieu equation, if the AD field is rotating
in a roughly circular orbit. That is to say,
the narrow-band  resonance occurs around 
$k/m_\Phi =  l$, $(l=1,2,\dots)$, and  $\delta \varphi$ grows 
unless the flat direction is moving in an exact circular orbit (i.e., $q=1-\varepsilon=0$).
The instability chart of Eq.~(\ref{eq:fluc_eq}) is illustrated in Fig.~\ref{fig:ins_chart}, and
one can see that the narrow-band structure around $\varepsilon = 1$ 
resembles those of  the Mathieu equation.

\begin{figure}
\includegraphics[width=7.5cm]{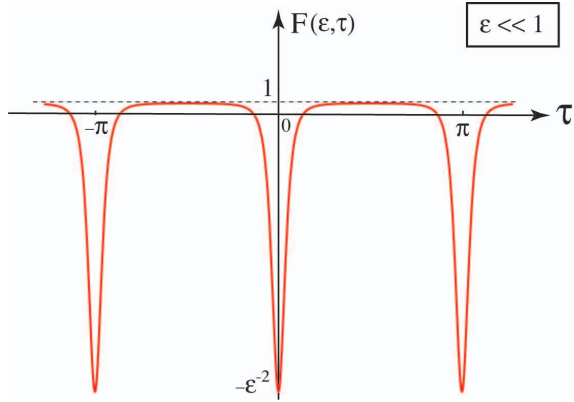}
\caption{
The typical behavior of $F(\varepsilon,\tau)$ when $\varepsilon \ll 1$.
}
\label{fig:F}
\end{figure}

Next we take up the other limit of $\varepsilon \ll 1$. In this case, the function $F(\varepsilon,\tau)$
behaves as follows. For $|\tau| \lesssim \varepsilon^{-1/2}$, it becomes negative and
the minimum locates at $\tau=0$: $F(\varepsilon,0)=- \varepsilon^{-2}$, while
$F(\varepsilon,\tau) \simeq 1$ for $|\tau| \gtrsim \varepsilon^{-1/2}$. See Fig.~\ref{fig:F}.
Therefore those fluctuations with $k < \varepsilon^{-1} m_{\Phi}$ experience tachyonic
instabilities and grow exponentially each time $\tau$ passes 
$\tau=n \pi,~(n=0, 1, 2, \dots)$. That is to say, the instability shows the step-function-like
behavior, as expected.
As can be seen from Fig.~\ref{fig:ins_chart},
the instabilities become stronger for smaller $\varepsilon$. However we expect that the
exponential growth rate will converge to some finite value $\sim O(1)$ as $\varepsilon 
\rightarrow 0$.

\begin{figure}
\includegraphics[width=7.5cm]{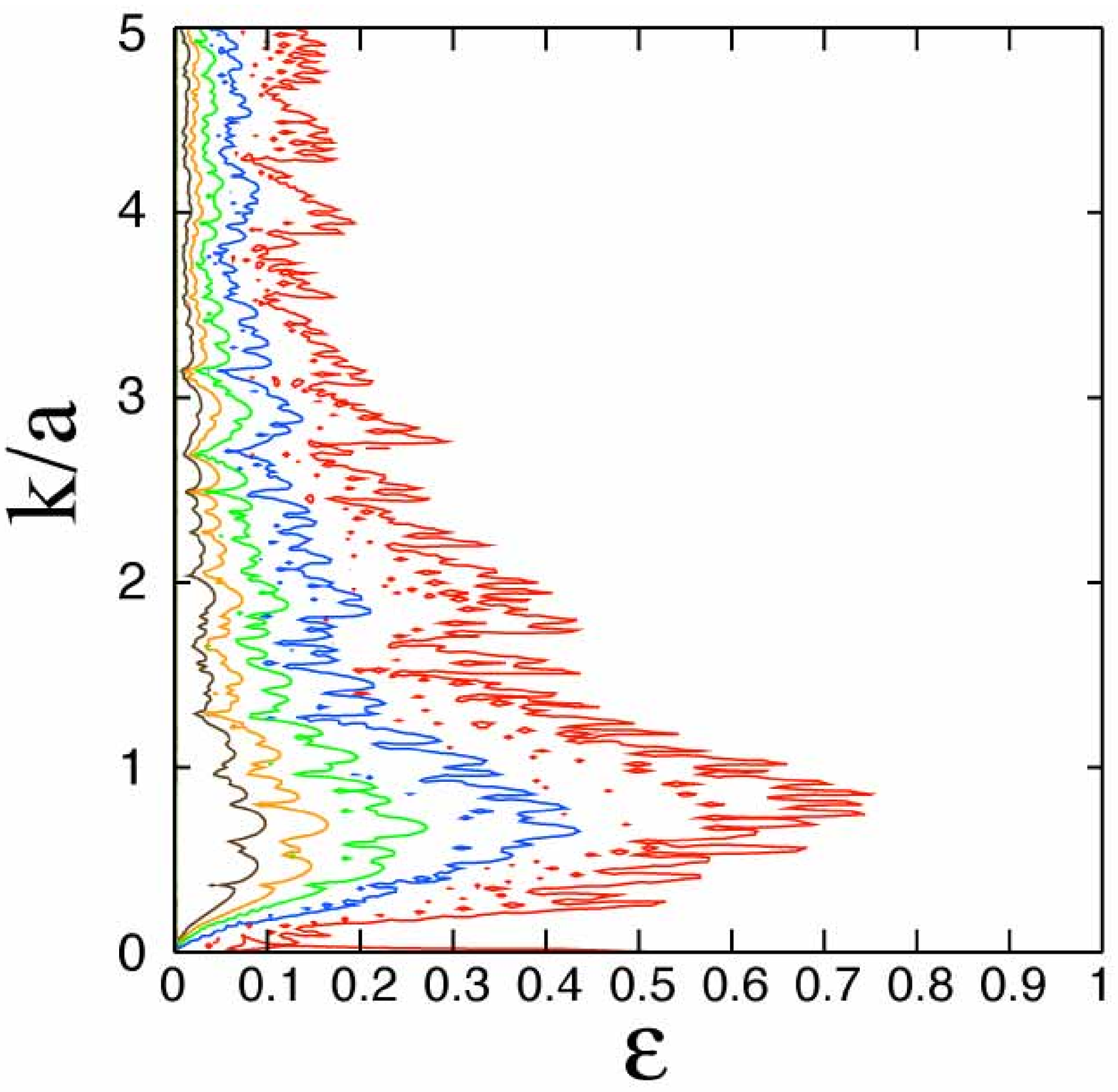}
\includegraphics[width=7.5cm]{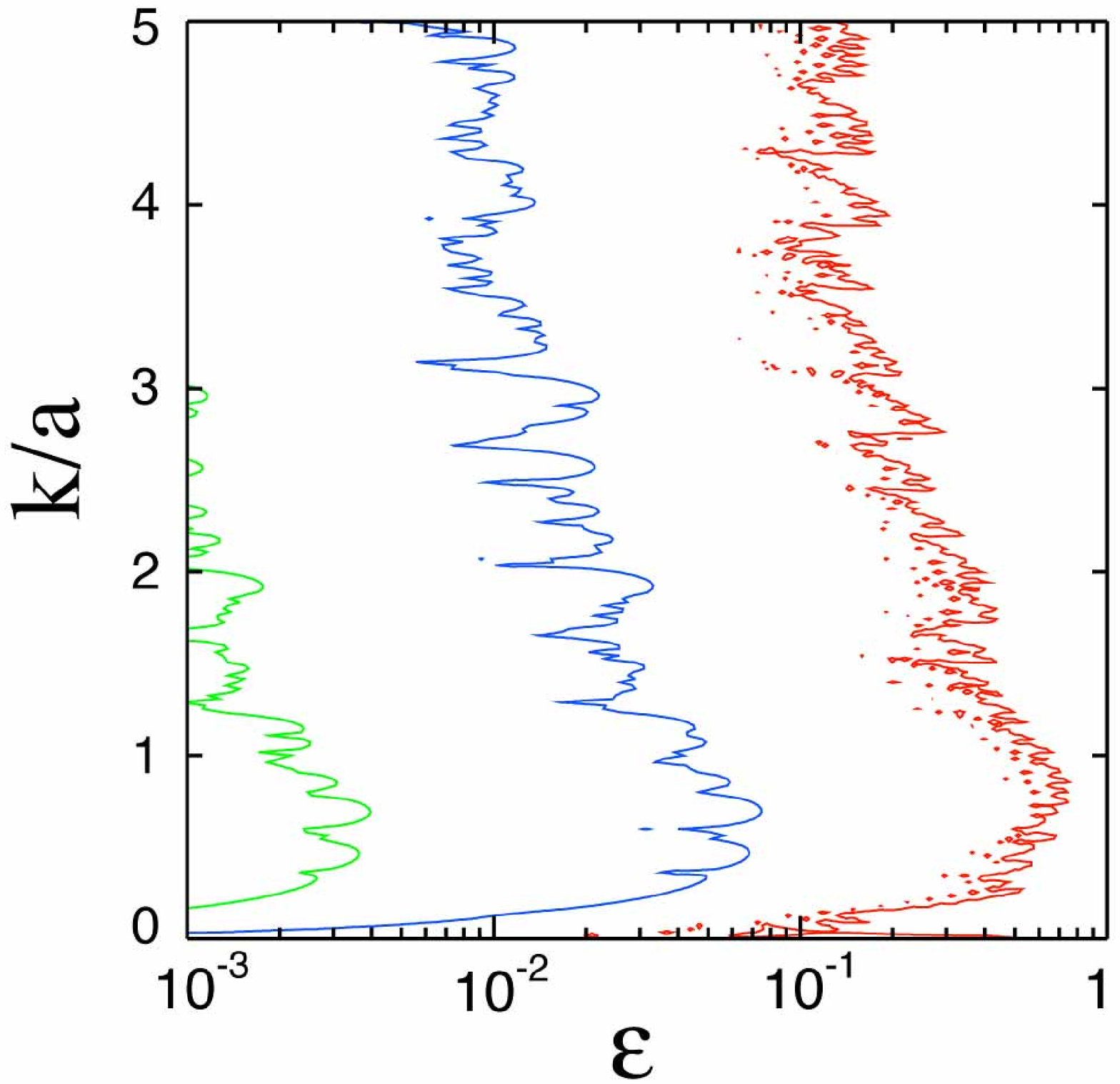}
\caption{
Same as Fig.~\ref{fig:ins_chart} except that the cosmic expansion
is taken into account. We take the initial condition as $H=m_\Phi$,
and assumed the matted-dominated like universe.
}
\label{fig:ins_chart_H}
\end{figure}

So far we have neglected the cosmic expansion, but the above-mentioned nature
of the instability does not change much even if the cosmic expansion is 
taken into consideration.  For simplicity, let us assume that the universe is dominated
by the energy of the oscillating inflaton, when the AD field starts to oscillate (i.e., $H\simeq m_{\Phi}$).
Then similar arguments lead to the equation of motion of $\delta \phi_m$ given by
Eq.~(\ref{eq:del_phim}) with $k^2$ replaced with $k^2/a^2$, where $a$ denotes the scale
factor. Note that the solution of the homogeneous mode is now given by
\bea
\phi_R = \phi_o a^{-3/2} \cos{m_\Phi t},\non\\
\phi_I = \varepsilon \phi_o a^{-3/2} \sin{m_\Phi t}.
\eea
Due to the redshift of the momentum $k$, the band structure becomes
vague as seen in Fig.~\ref{fig:ins_chart_H}.  However, except for
$\varepsilon \sim 1$ where the instability is  narrow-band resonance,
there still exist strong $D$-term instabilities because of their tachyonic nature.

The above argument is not limited to the parabolic potential, but it 
can easily be generalized to an arbitrary potential with $U(1)$ symmetry.
Although the collective fluctuations as well are subject to instabilities peculiar
to the potential in general, we can concentrate on the relative fluctuations by adopting
Eq.~(\ref{eq:rel_app}) since they are independent.
As long as the homogeneous motion satisfies the $D$-flat condition,
the lighter one of the relative fluctuations satisfies
\beq
\label{eq:del_phim_gene}
\delta \ddot{\phi}_{m}+2\frac{\dot{\phi}}{\phi}\delta \dot{\phi}_m+ k^2 \delta \phi_m = 0,
\eeq
where the definitions of the variables are same as before. If we rewrite this
equation in terms of $\delta \varphi \equiv \phi \,\delta \phi_m$, we have
\beq
\label{eq:mathieu_gene}
\delta \ddot{ \varphi}+\left(k^2-\frac{\ddot{\phi}}{\phi}\right) \delta \varphi=0.
\eeq
When the AD field comes the closest to the origin, $\ddot{\phi}/\phi$ is
expected to take its maximum value, which leads to the tachyonic
instabilities for $k^2 < [\ddot{\phi}/\phi]_{\rm max}$.  Also if the homogeneous
motion in the limit of $\varepsilon \rightarrow 1$ can be approximated as 
\bea
\label{eq:axial_gene}
\phi_R = \phi_o \cos{\Omega t},\non\\
\phi_I = \varepsilon \phi_o \sin{\Omega t},
\eea
then Eq.~(\ref{eq:mathieu_gene}) reduces to the Mathieu equation with $A=k^2$ and
$q \simeq \Omega^2 (1-\varepsilon)$. In this case,
the instability bands are narrow and locate at $k/\Omega = l,~(l=1,2,\dots)$.
Thus the existence of the $D$-term instability does not depend on the specific form of
the potential.

In this section we have considered a simple model of $D$-flat directions. However the
existence of the instability should be common to all $D$-flat directions. The reason
is as follows. First, all the MSSM flat direction is at least
charged under $U(1)_Y$. Thus $D$-term potential similar to Eq.~(\ref{eq:potential12}) 
always exists. Second,
the instability originates from a massless mode of the relative fluctuations between 
the constituent fields, the existence of which is a common feature of $D$-flat directions.
To sum up the major characteristics of the $D$-term
instability, (i) for $\varepsilon \sim 1$, it is narrow-band resonance similar to the
one expresssed by 
the Mathieu equation; 
(ii) the tachyonic instability appears for $\varepsilon < 1$; (iii) it is 
 independent of amplitude of the flat direction. 
In the following sections we will see how such $D$-term instability 
affects the dynamics of the AD field.


\section{Effects on $Q$-ball formation} 
\label{sec:qb_form}
\subsection{Review of $Q$-balls and $I$-balls} 
\label{sec:qb_ib}

Let us now discuss how the $D$-term instability affects 
the $Q$-ball formation processes. Before we go into details, it will be useful to
briefly review the $Q$-ball formation. For the moment we adopt the single-field
parametrization. Let a flat direction $\Phi$ possess an unit baryon charge.
During inflation $\Phi$ is assumed to take a large
expectation value. When the Hubble
parameter  becomes comparable to the mass of
$\Phi$ after inflation ends, $\Phi$ starts to oscillate in the potential
 Eq.~(\ref{eq:phi_pote}). If the sign of $K$ is
negative, it experiences spatial instabilities, leading to the
$Q$-ball formation~\cite{Kusenko:1997si,Enqvist:1998en,Kasuya:1999wu}. 
The profile of thus formed gravity-mediation type $Q$-balls
 is approximated by the spherically symmetric gaussian form with 
good precision:~\cite{Enqvist:1998en}
\bea
\Phi &=& \frac{\phi_R+i \phi_I}{\sqrt{2}},\non\\
\phi_R(r,t)&=&\phi_o \,e^{-r^2/R_Q^2} \cos{\omega t} ,\non\\
\phi_I(r,t)&=&\phi_o \,e^{-r^2/R_Q^2} \sin{\omega t}, 
\eea
with 
\bea
\omega &\equiv& m_\Phi \sqrt{1+2|K|},\non\\
R_Q &\equiv& \frac{\sqrt{2}}{\sqrt{|K|} m_\Phi},
\eea
where $\omega$ and $R_Q$ denote the angular velocity and radius of the $Q$-ball,
respectively.

Although a $Q$-ball solution is realized as the energy minimum state with a fixed
global $U(1)$ charge, it is a  $I$-ball that is first formed in general. The profile of $I$-balls
is almost same as that of $Q$-balls except for axial ratio of the orbit:
\bea
\phi_R(r,t)&=&\phi_o \,e^{-r^2/R_Q^2} \cos{\omega t} ,\non\\
\phi_I(r,t)&=&\varepsilon_I \,\phi_o \,e^{-r^2/R_Q^2} \sin{\omega t}, ~~~(0 \leq \varepsilon_I < 1).
\eea
The reason why $I$-balls are formed instead of $Q$-balls is that
the quadratic potential Eq.~(\ref{eq:phi_pote}) allows the dynamics of system
to have another invariant, the adiabatic charge $I$~\cite{Kasuya:2002zs},
and that the axial ratio of the homogeneous motion is generally less than $1$.
Thanks to the conservation of the adiabatic charge, $I$-balls are formed 
as a quasi-stable state, and the axial ratio of the orbit inside $I$-balls, $\varepsilon_I$,
is almost  same as that of homogeneous mode before fluctuations grow, $\varepsilon_H$:
$$
\varepsilon_I \simeq \varepsilon_H.
$$
However this relation does not hold when the $D$-term instability is taken into account,
and $\varepsilon_I $  is larger than or equal to $\varepsilon_H$ as shown below.

Central to this issue is how $I$-balls eventually transform into $Q$-balls.
Due to the elliptical orbit, $I$-balls are considered to be an excited state of $Q$-balls.
However it was unclear to date how $I$-balls emit the extra energy to settle
down to $Q$-balls, since they have very long lifetime, which makes it difficult
to follow the evolution in numerical calculations. Here we would like to argue that
the $D$-term instability violates the conservation of the adiabatic charge, which 
enables $I$-balls, if formed,  to immediately transform into $Q$-balls
by emitting both the energy and $U(1)$ charge.

Before we proceed, however, it is necessary to ask whether $I$-balls are 
really formed under the influence of the $D$-term instabilities. 
Since the exponential growth rate of the $I$-ball
formation is given by $\mu_Q \sim |K| m_{\Phi} \sim 0.1 m_{\Phi}$~\cite{Enqvist:1998en},
the $I$-ball formation will complete before the $D$-term instabilities grow sufficiently,
if the initial axial ratio $\varepsilon_{H}$ is larger than $\sim 0.7$ (see Fig.~\ref{fig:ins_chart_H}).
Furthermore, even if the $D$-term instabilities grows faster than $\sim |K| m$, 
some $I$-ball-like objects might be generated as a result of the interactions between
the developed fluctuations, because
the fastest growing mode of the $D$-term instability locates around $k/a \sim m_\Phi$
which is not so different from the typical size of $I$-balls. 
 In fact, we have performed numerical calculations not taking account of
 the cosmic expansion and confirmed that
 $I$-balls  are formed for  
 $\varepsilon_{H} \geq 0.1$, but the axial ratio of the orbit inside 
 thus formed lumps is larger
 than $\varepsilon_{H}$ from the beginning, e.g., $\varepsilon_I\sim 0.3$ for 
 $\varepsilon_{H} = 0.1$.  In particular, $I$-balls with $\varepsilon_I \lesssim
 0.2$ are not formed, probably because the $D$-term instability with $\varepsilon_I \lesssim
 0.2$ is too strong for $I$-balls to retain their configuration.  Although the cosmic
 expansion might make the interactions between the fluctuations less effective,
some part of the charge asymmetry will still be stored inside such $I$-ball-like
objects.  We will discuss the evolution of the axial ratio of the trajectory inside these objects 
in the next two subsections.

The $I$-ball formation is not the end of the story. Once $I$-balls are formed,
the dynamics of the AD field inside them decouples from the cosmic expansion.
Since the $D$-term instabilities lie at smaller scales than a typical size of  $I$-balls, 
fluctuations inside them continue to grow exponentially. It should be noted that
the $D$-term instability does not depend on the amplitude, so that they will grow  
throughout the $I$-ball.  The growth  will last either
until the conversion of $I$-balls into $Q$-balls completes (i.e., the axial ratio $\varepsilon$
becomes 1), or until $I$-balls are completely destroyed by the instability.  We need to
resort to numerical calculations to determine which is realized.

\subsection{Numerical Calculations} 
\label{sec:num}
As mentioned in the previous subsection, we have numerically investigated  the
formation processes of $I$-balls taking account of the $D$-term instability and
found that  almost all the charge asymmetry is taken in the $I$-balls for
$\varepsilon_{H,{\rm ini}} \geq 0.1$:
\beq
\label{eq:fi}
f_I \simeq 1,
\eeq
where $f_I$ denotes the fraction of the charge absorbed by the $I$-balls. Due to
the technical reasons, it is difficult both to take account of the cosmic expansion
and to follow the evolution of the AD field with $\varepsilon_{H,{\rm ini}} \leq 0.1$.
The cosmic expansion might change the $I$-ball formation inefficient to some extent.
The strong instability for $\varepsilon_{H,{\rm ini}} \ll 0.1$ as well  prevents the $I$-balls
to be formed. But what we would like to show here is that the charge asymmetry is partly
left behind during the $Q$-ball formation, even if it is once absorbed by $I$-balls.

\begin{figure}
\includegraphics[width=7.5cm]{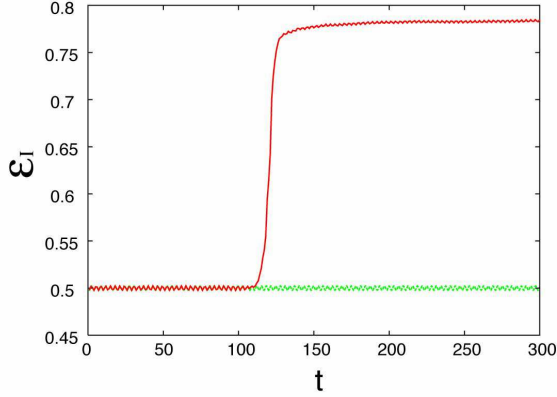}
\caption{
Evolutions of the  axial ratio of the orbit inside a $I$-ball
with the initial value:$\varepsilon_I=0.5$. Due to the $D$-term
instability, $\varepsilon_I$ tends to approach $1$ (solid line),
while it remains constant 
in the single-field parametrization (dotted line). 
}
\label{fig:epsilon}
\end{figure}

\begin{figure}
\includegraphics[width=7.5cm]{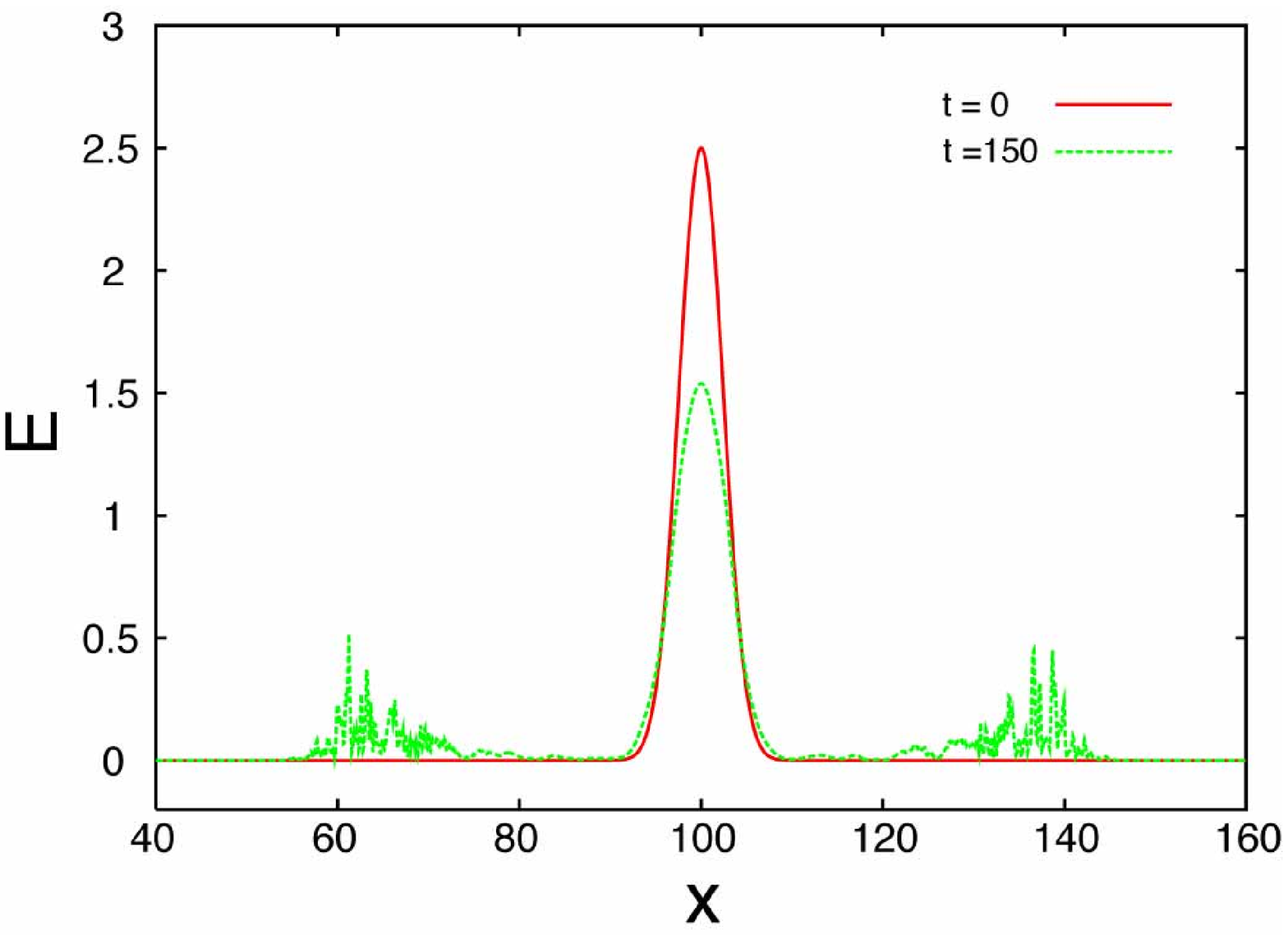}
\includegraphics[width=7.5cm]{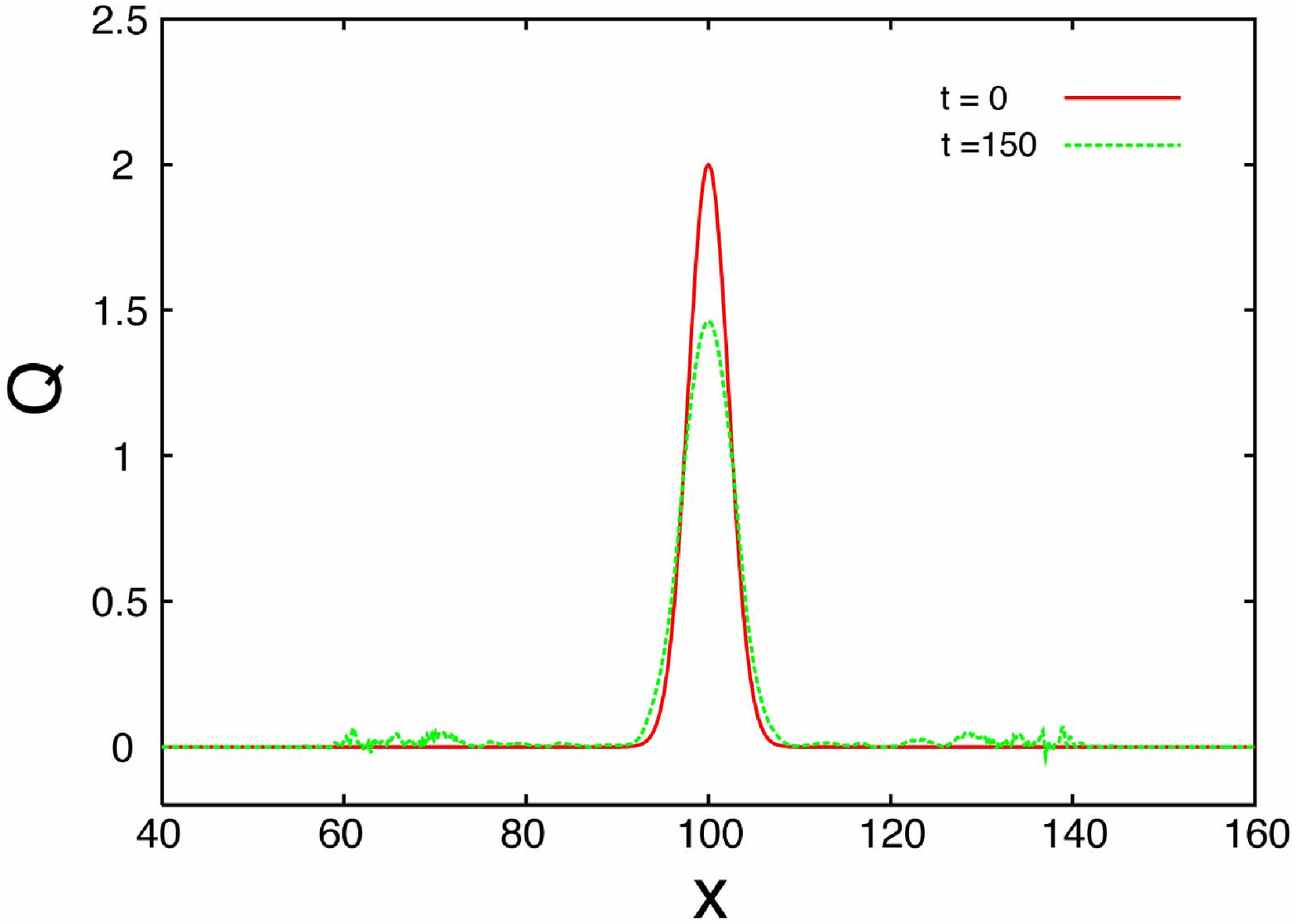}
\caption{
Spatial distributions of energy and charge of the $I$-ball 
with the initial axial ratio: $\varepsilon_I = 0.5$. The $I$-ball becomes more $Q$-ball-like by emitting both the energy and the charge. Note that the energy is more efficiently
released than the charge, which makes the effective axial ratio increase.
}
\label{fig:E_Q}
\end{figure}

Assuming that  $I$-balls are formed, we have followed the evolution of the $I$-ball
in order to show that the $D$-term instability actually occurs inside $I$-balls. 
We have solved the equations of motion given by
Eq.~(\ref{eq:EOM}) on the one dimensional lattices with the initial condition:
\bea
\phi_{1R}=\phi_{2R}&=&\phi_o \,e^{-r^2/R_Q^2} \cos{m_\Phi t} ,\non\\
\phi_{1I}=\phi_{2I}
&=&\varepsilon_I \,\phi_o \,e^{-r^2/R_Q^2} \sin{m_\Phi t}.
\eea
Once $I$-balls are formed, the dynamics inside them decouples
from the cosmic expansion. Therefore we can safely neglect the effect 
in contrary to the case before the $I$-balls are formed. It should be reminded 
that the axial ratio $\varepsilon_I$  in the above  equation does not necessarily coincides 
with that of the homogeneous motion $\varepsilon_H$.
The space $x$ and time $t$ are normalized by
the mass scale $m_\Phi$, while the field value is normalized by $M_*$. We
take the following values for the model parameters: $m_1=m_2=m_\Phi$,
$\phi_o/M_*=1$, $g^2 M_*^2 = 10^3~{\rm or~} 10^4\, m_{\Phi}^2$, and $K_1=K_2=-0.1$.
The typical evolution of the  axial ratio $\varepsilon_{I}$ with the
initial value: $\varepsilon_I=0.5$, is illustrated in Fig.~\ref{fig:epsilon}. 
The spatial distributions of the energy and the charge are shown in
Fig.~\ref{fig:E_Q}.  As seen in the figure, the axial ratio increases, but
it does not reach $1$ due to the following reasons. 
First of all, as the axial ratio comes closer to $1$, the instability becomes
weaker as shown in Fig.~\ref{fig:ins_chart}. 
Thus in the finite CPU time the complete conversion of the $I$-ball
to a $Q$-ball cannot be seen. Second, the finite lattice size allows the decay 
product emitted from the $I$-ball to be absorbed by the $I$-ball again. 
This effect as well prevents the $I$-ball to transform into the $Q$-ball.
(We have chosen the time interval in Fig.~\ref{fig:epsilon}, so that this effect is  
negligible.) However, in the real world, such a tendency toward $Q$-balls
will last throughout,  because the adiabatic charge is no longer conserved
due to the $D$-term instability.
We thus expect the conversion will be completed relatively soon 
after the $I$-ball formation.

\begin{figure}
\includegraphics[width=7.5cm]{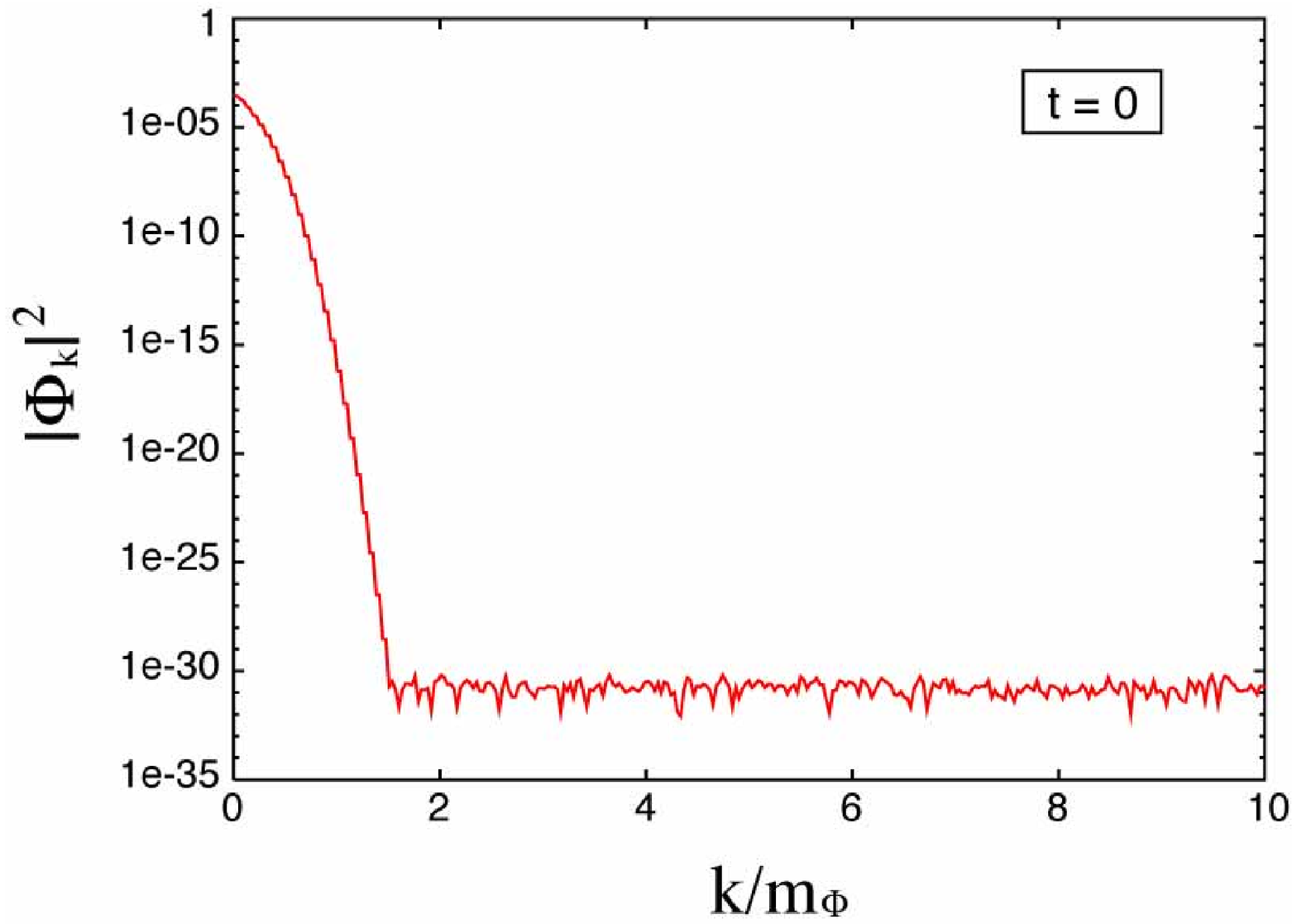}
\includegraphics[width=7.5cm]{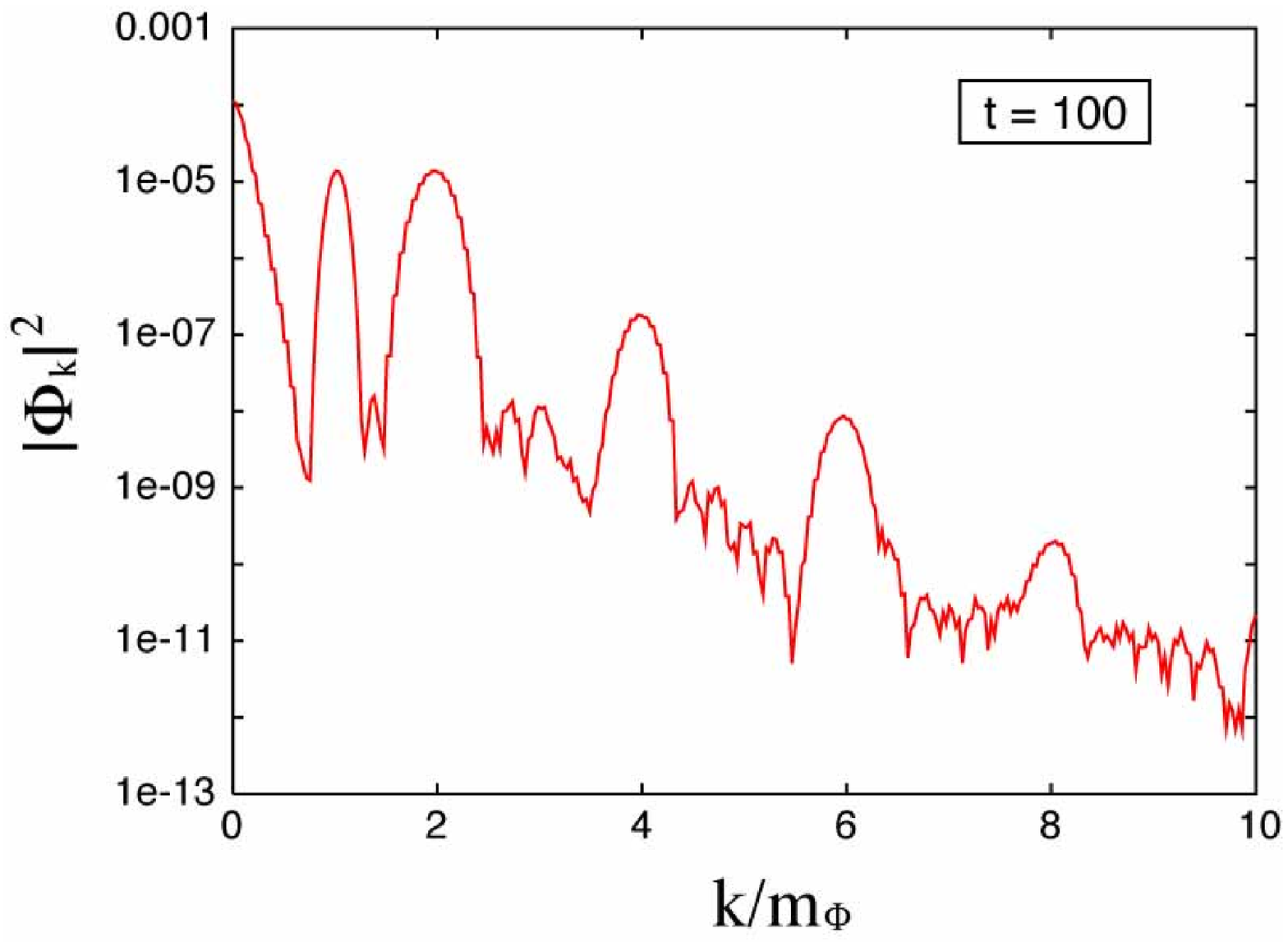}
\caption{
Power spectra of the constituent field $\Phi_1$.
The model parameters are same as before except $K_1=K_2=-0.01$.
The upper panel represents the power spectrum of the initial $I$-ball
solution with $\varepsilon_I=0.5$,
 while the $D$-term instability begins to grow in the lower panel.
The instability bands coincide with those obtained by the analysis of
linearized fluctuations (see Fig.~\ref{fig:ins_chart}).
The instability bands corresponding to $k/m_\Phi = 2, 4, 6,\dots$
are enhanced due to``rescattering" processes.
}
\label{fig:fft}
\end{figure}

We have also calculated the power spectra of one of the constituent fields, $\Phi_1$. 
See Fig.~\ref{fig:fft}. 
We take the size of the $I$-ball
larger (i.e., smaller $|K|$) to obtain clear signature of the  instability band. 
The result coincides with the instability band obtained
by the analysis of linearized fluctuations.  Thus we conclude that the $D$-term
instability plays an essential role to transform $I$-balls into $Q$-balls.

\subsection{Modified relation between baryon and dark matter abundance} 
\label{sec:b_dm}
What should be emphasized  at this juncture 
is that both the energy and charge are emitted from the $I$-ball.
This is because the four-point interaction in the $D$-term inevitably induces
both the charge-transferring (e.g., $\Phi_1 \Phi_1 \rightarrow \Phi_1\Phi_1$)
and charge-conserving processes (e.g., $\Phi_1 \Phi_1^* \rightarrow \Phi_1\Phi_1^*$). 
If almost all the charge is absorbed by the $I$-balls at their formation,
the fraction of the baryon charge stored inside $Q$-balls, $f_Q$, 
depends only on  the initial axial ratio~\footnote{
Here we assume that the charge emitted from $I$-balls
due to the $D$-term instability dominate over that evaporated 
from the surface of $Q$-balls
through the interaction with thermal plasma. In fact, the emitted charge can be
as large as that of $Q$-balls due to the $D$-term instability.
}.
Since $f_Q$ is defined as the fraction of the charge stored inside $Q$-balls
assuming all the charge asymmetry is once absorbed by $I$-balls, the actual
fraction kept inside $Q$-balls is $f_B\equiv f_I f_Q$.
It is important to know the value of $f_B$ to relate the baryon number to
dark matter density. 
The numerical results of $f_Q$ for several values of $\varepsilon_I$
are summarized in Table~\ref{tab:fb}.
The values of $f_Q$ shown in the table should be considered to
give upper bounds of $f_B$, since 
some part of the charge is left behind at the $I$-ball formation and 
also the conversion is not yet completed. 
It should be noted that the $I$-ball comes apart for $\varepsilon_I \leq 0.2$ due to the strong
$D$-term instability. This fact agrees with the  result that $I$-balls with 
the  axial ratio smaller than $\sim0.2$ are not formed.

The neutralino LSPs produced by the decay of $Q$-balls survive and can be
the dark matter, if  the decay temperature of $Q$-balls is
lower than the freeze-out temperature of the LSP. Let $N$ be  the average number of the LSPs produced by the
decay of unit baryon charge of $Q$-balls. The abundance of the LSP dark matter is then 
related to that of baryon number as follows.
\beq
\Omega_{DM} = 3 \left(\frac{N}{3}\right) \left( \frac{m_{\rm LSP}}{m_n} \right)
 f_B \Omega_B,
\eeq
where $\Omega_{DM, \,B}$ denote the density parameters of the dark matter
and the baryon, $m_{\rm LSP}$ and $m_n$ are masses of the LSP
and  the nucleon, respectively. Substituting the WMAP results~\cite{Spergel:2003cb}: $\Omega_{DM} h^2= 0.116$ and $\Omega_B h^2= 0.024$, we obtain
\beq
f_B \sim 0.02 \left(\frac{3}{N}\right) \left(\frac{m_{\rm LSP}}{100 {\rm GeV}} \right)^{-1}
\eeq
Previously, $f_B \sim 1$ was supported by the numerical 
calculations~\cite{Kasuya:2000wx}, therefore
the coincidence of the baryon and dark matter abundance cannot  be explained in this way.
However, thanks to the $D$-term instability, $f_I \ll 1$ might be achieved if $\varepsilon_H$ is
much less than $\sim 0.1$. 
Since the flat direction completely decays before $I$-balls are formed 
for small enough $\varepsilon_H$, there must be an intermediate value of $\varepsilon_H$ that
leads to $f_B \sim 0.02$.

On the other hand,  the new-type $Q$-balls~\cite{Kasuya:2000sc} themselves
can be the dark matter in the gauge-mediated SUSY breaking models.
Since the new-type $Q$-balls are formed where the gravity-mediated
potential overwhelms the gauge-mediated one, the above-mentioned
arguments can be applied. If the energy per unit charge of the new-type
$Q$-ball is smaller than the nucleon mass, they are stable and can
be the dark matter. The relation between the dark matter and the baryon
abundance is thus given by
\beq
\Omega_{DM} = \frac{f_B}{1-f_B} \frac{m_{3/2}}{m_n} \Omega_B,
\eeq
where we take $m_\Phi \simeq m_{3/2}$. This gives the constraint
on $f_B$:
\bea
f_B &=& \frac{\Omega_{DM}}{\Omega_{DM}+\frac{m_{3/2}}{m_n} \Omega_B},\non\\
      & \simeq&1.0,~ 0.98,~0.83,~~~~{\rm for~~}m_{3/2}=0.01,~0.1,~1 {\rm GeV},
\eea
which can be explained by taking e.g., $\varepsilon_H \gtrsim 0.4$ if $f_I \simeq 1$ and
$\varepsilon_H \simeq \varepsilon_I$.

It has been thought that almost all the baryon number is
absorbed in $Q$-balls.  That is why it was necessary to estimate
how much the baryon charge evaporates from the surface of $Q$-balls
through the interaction with thermal plasma. However, in fact, some finite
fraction is emitted from $I$-balls due to the $D$-term instability, and 
the fraction is determined only by
the initial axial ratio, independent of the reheating temperature and so on.
Thus the $D$-term instability may revive the appealing solution for the
coincidence problem between the dark matter and baryon abundance,
which was otherwise difficult to be achieved.


\begin{table}
\caption{Numerical results of the final axial ratio and $f_Q$. 
For $\varepsilon_I \leq 0.2$, the $I$-ball divides into mutliple $Q$-balls,
and the final axial ratio and $f_Q$ of the largest one are shown in parenthesis below.
For $\varepsilon_I=0.9$, the conversion of the $I$-ball to a $Q$-ball cannot be observed
in the numerical calculation. 
}
\begin{center}
\begin{tabular}{c||c|c|c|c|c|c|c|c|c}
$\varepsilon_{I,{\rm ini}}$&0.1&0.2&0.3&0.4&0.5&0.6&0.7&0.8&0.9 \\
\hline
$\varepsilon_{I,{\rm fin}}$&(0.5)&(0.6)&0.7&0.73&0.78&0.8&0.82&0.85&0.9 \\
\hline
$f_Q$&(0.15)&(0.6)&0.72&0.85&0.93&0.96&0.97&0.99&1.0
\end{tabular}
\end{center}
\label{tab:fb}
\end{table}
%

\section{Fate of flat directions} 
\label{sec:tachyonic}
In the previous section we have considered the case that
$Q$-balls are formed. What if $Q$-balls are not formed?
If the coefficient of the one-loop correction $K$ is positive,
 $Q$-balls are not formed. Also even if $K$ is negative,
the strong tachyonic instabilities will prevent $Q$-balls from
being formed, when the initial axial ratio is small enough.
It should be noted that $Q$-balls remain stable under the
effect of the $D$-term instability since the instability disappears 
when the orbit becomes circular. In other words, unless $Q$-balls
are formed, the $D$-term instability will never cease to grow.
Therefore, if $K<0$ or $K>0$ with $\varepsilon_H \ll 1$, 
the flat direction experiences the  $D$-term instability until it
deforms into a disordered state, which will decay and/or interact with other
particles and attain the thermal equilibrium sooner or later. 
The fate of the flat directions is thus very simple: $Q$-ball or nothing.

From the above discussion, the gravity-mediation type $Q$-balls are
formed  when the initial axial ratio of the orbit is large enough 
$\varepsilon_H \gtrsim 0.1$. Then what about the gauge-mediation type
$Q$-balls?  Since the strength of the $A$-term is proportional to the
gravitino mass, the initial axial ratio tends to be much smaller than $1$
if the AD field start to oscillate in the gauge-mediated 
potential~\cite{deGouvea:1997tn}.
Since the homogeneous motion in the gauge-mediated potential 
cannot be solved analytically, it is necessary
to perform numerical calculations to investigate  the $D$-term 
instabilities. Due to this additional instability, the $Q$-ball formation will be considerably
delayed or prevented. The detailed discussion on this subject
will be presented elsewhere~\cite{preparation}.

\section{discussions} 
\label{sec:discuss}
In this paper we have investigated the dynamics of the flat directions 
taking account of the $D$-term potential. To this end, we adopted
the multi-field parametrization of the flat directions, which enables us
to deal with the relative fluctuations between the constituent fields.  
It is found that there exist instabilities due to the 
D-term potential and the nature of these instabilities depends
on the axial ratio $\varepsilon$ of the oscillating flat direction.
If $\varepsilon$ is close to $1$, the $D$-term instability is
 similar to narrow resonance described by the Mathieu equation
with $q \ll 1$. On the other hand, the instabilities show the
tachyonic nature for $\varepsilon <1$. Due to this tachyonic instability,
the cosmological evolution of the flat direction are always under
the influence of $D$-term instability.
In particular we have shown that the existence of the instabilities is crucial 
to the formation of $Q$-balls. Since the $D$-term instability violates the 
conservation of the adiabatic charge,
$I$-balls are no longer quasi-stable, and therefore 
the transition from $I$-balls to $Q$-balls occurs very efficiently.
It is important to note that  during this process some fraction
of the charge of the $Q$-ball is emitted, which
 may revive the relation between the baryon 
number and dark matter density of the universe. 
Furthermore, the $D$-term instability drastically changes the
evolution of the flat directions. Unless $Q$-balls are formed, the
flat direction completely decays relatively soon after it begins to oscillate.
We believe that the existence of such instabilities is generic to the system
of scalar fields with the $D$-term potential. For instance, in the reheating stage
of the $D$-term inflation models~\cite{Binetruy:1996xj}, the scalar fields are
oscillating in the similar potential. Thus the reheating processes of inflation models
of this type might be affected by the $D$-term instability. However, further investigation
is clearly necessary and is beyond the scope of this paper.

\subsection*{ACKNOWLEDGMENTS}
We would like to thank S. Kasuya for reading the manuscript and useful comments. 
F.T. would like to thank the Japan Society for the Promotion
of Science for financial support. 
This work was partially supported by the JSPS Grant-in-Aid for Scientific 
Research  No.\ 14540245 (M.K.).

 \end{document}